\documentclass[aps,pre,showpacs,onecolumn,superscriptaddress,11p]{revtex4}
\usepackage{epsfig}
\usepackage{amsmath,amssymb,epsfig,capt-of,ifthen,calc}
\usepackage{latexsym}   
\usepackage{setspace}   
\usepackage{array}      
\usepackage{delarray}   
\usepackage{afterpage}
\usepackage{graphicx}
\usepackage{dcolumn}
\usepackage{bm}
\usepackage{array}
\usepackage{hyperref}
\usepackage{float}
\usepackage{supertabular}
\usepackage{longtable}
\newcommand{\be}{\begin{equation}}
\newcommand{\ee}{\end{equation}}
\newcommand{\bea}{\begin{eqnarray}}
\newcommand{\eea}{\end{eqnarray}}

\newcommand{\vishal}[1]{{\large{$\clubsuit$}}{\em #1}}
\newcommand{\orion}[1]{{\large{$\spadesuit$}}{\em #1}}

\usepackage{amsmath}

\begin{document}
\bibliographystyle{unsrt}

\title{Node Similarity Within Subgraphs of Protein Interaction Networks}

\author{Orion Penner} 
\affiliation{Complexity Science Group,
  University of Calgary, Calgary, Alberta T2N 1N4, Canada}
\author{Vishal Sood} 
\affiliation{Complexity Science Group,
  University of Calgary, Calgary, Alberta T2N 1N4, Canada} \affiliation{Institute for
  Biocomplexity and Informatics, University of Calgary, Calgary, Alberta T2N 1N4,
  Canada} 
\author{Gabriel Musso} 
\affiliation{Department of Medical Genetics and Microbiology, 
University of Toronto, Toronto, Ontario M5S 3E1, Canada}
\author{Kim Baskerville} 
\affiliation{Perimeter Institute for Theoretical Physics,
  Waterloo, Ontario N2L 2Y5, Canada}
\author{Peter Grassberger} \affiliation{Complexity Science Group,
  University of Calgary, Calgary, Alberta T2N 1N4, Canada} 
\affiliation{Institute for
  Biocomplexity and Informatics, University of Calgary, Calgary, Alberta T2N 1N4,
  Canada} 
\author{Maya Paczuski} \affiliation{Complexity Science Group,
  University of Calgary, Calgary, Alberta T2N 1N4, Canada}


\date{\today}

\begin{abstract}
  We propose a biologically motivated quantity,
  {\it twinness}, to evaluate local similarity between nodes in a network.
  The twinness of a pair of nodes is the  number of connected, labeled subgraphs of size $n$ in which the two nodes possess identical neighbours. 
  The graph animal algorithm  is used
to estimate twinness for each pair of nodes (for subgraph sizes $n=4$ to $n=12$) in four different protein
  interaction networks (PINs).  These include an {\it  Escherichia coli} PIN and three {\it
    Saccharomyces cerevisiae} PINs -- each obtained using state-of-the-art
 high throughput methods. In almost all cases, the
  average twinness of node pairs is vastly higher
  than expected from a null model obtained by switching links. For all $n$, we observe a difference in the ratio of type ${\bf A}$ twins (which are
{\it unlinked} pairs) to type ${\bf B}$ twins (which are {\it linked} pairs)
 distinguishing the prokaryote {\it  E.~coli}  from the eukaryote
  {\it  S.~cerevisiae}. 
  Interaction similarity is
  expected due to gene duplication, and whole genome duplication paralogues
  in {\it S.~cerevisiae}
  have been reported to co-cluster into the same complexes.
  Indeed, we find that these paralogous proteins
  are  over-represented as twins compared to pairs  chosen at random. These results indicate that 
  twinness can detect ancestral relationships from currently available PIN
  data.

\end{abstract}

\maketitle

\section{Introduction}
Proteins constitute the machinery that carry out cellular processes by
forming stable or transitory complexes with each other --
organized perhaps into a
web of overlapping modules.
Information about this complex system  can be
condensed into a protein interaction network (PIN), which is a
graph where nodes are proteins and links are
measured or inferred pairwise binding interactions in a cell.  Major efforts
over the years devoted to resolving protein interactions have employed both
small-scale and large-scale techniques. High throughput methods, such
as  yeast two hybrid (Y2H) and  tandem affinity purification
(TAP) have recently generated vast amounts of protein interaction
data~\cite{butland::nature2005,krogan::nature2006,gavin::nature2006}, allowing
PINs from different organisms, experiments, research teams etc. to be compared.

\begin{figure}
\centering
\includegraphics[width=0.7\textwidth]{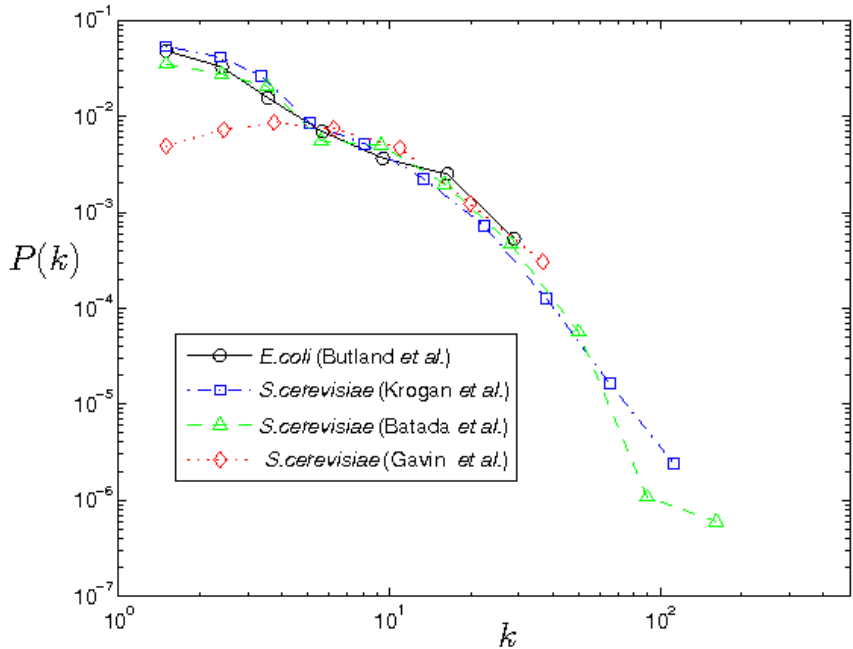}
\caption{Degree distribution $P(k)$ as a function of degree $k$ for four
different PINs. Note the systematic curvature and the variation between
different data sets for the budding yeast {\it S.~cerevisiae}.\label{fig:degdist}}
\end{figure}

A basic statistical feature of any network is its degree distribution, $P(k)$, for
the number of links, $k$, connected to a node. In this respect,
a variety of networks have been shown~{\cite{barabasi::rmp2002}} to deviate decisively from a
random graph, where the degree distribution is Poisson.  In fact, early work suggested that degree distributions for PINs were power-law or scale-free~\cite{barabasi::nature2001}.
However, as demonstrated in Fig.~\ref{fig:degdist}, degree
distributions for recently obtained PINs are neither power-law nor particularly stable
across different state-of-the-art constructions for the same organism -- here the
budding yeast {\it S.~cerevisiae}.  Note that all of the data sets studied here are based on the TAP-MS technique, 
except for Batada {\it et al.}, which is a compilation of data obtained from a number of different techniques.

Despite the empirical inconsistency presented by PIN degree distributions, similar local structures can stand out when each network is compared to a null model where links are switched while retaining the original degree sequence~\cite{alon::science2002,newman::siam2003}.  Subgraphs that are significantly over-abundant are referred to as motifs, while subgraphs that are significantly under-represented are referred to as anti-motifs~\cite{alon::science2002}.  It has been reported that proteins within motifs are more conserved than other proteins~{\cite{barabasi::natgen2003}}.  In PINs, dense subgraphs containing 3 or 4 nodes are motifs, while tree-like subgraphs are anti-motifs~\cite{baskerville::arxives2007}.

Complementary to the search for motifs, graph clustering algorithms have been applied to identify components or complexes in PINs. By construction, these components tend to contain a high density of links but are weakly connected to the rest of the network (see e.g. Refs.~\cite{spirin::pnas2003,wang::plos2007}).  Complexes identified in this manner can contain up to 100 or more proteins, with on-going debate~\cite{wang::plos2007} as to their biological significance. However, biological processes such as signal transduction, cell-fate regulation, transcription, and translation typically involve a few tens of proteins.  In previous work~\cite{spirin::pnas2003} mesoscale (5-25) protein clusters have also been identified using graph clustering algorithms. These clusters were matched with groups of proteins known to form complex macromolecular structures, or modules of proteins that participate in a specific function in the cell~\cite{spirin::pnas2003}.

The subgraphs we probe lie within the mesoscale range, but our technique avoids using clustering methods to identify components.  Rather we directly estimate frequencies of occurrence of connected subgraphs of different sizes and analyze properties of these subgraph collections in the network. The particular properties we focus on here involve node similarity within subgraphs, quantified through a measure we refer to as 'twinness'.

Just as degree distributions are not consistent across currently available PINs,
extensive studies of
subgraph counts~\cite{baskerville::arxives2007} as well as modular structures~\cite{wang::plos2007}
show discrepancies among PINs nominally representing the same
organism (again {\it S.~cerevisiae}). 
Hence, in cases where the results
are not demonstrably robust over current data sets,
one can expect conclusions based on these results to change as the methods
to construct PINs improve.
On the other hand, statistical measurements that are
relatively consistent over different PINs for the same 
organism can {\it a priori} be considered more robust.  This is the criteria
we use to decide whether or not results from our analyses are reliable.

A dominant feature of biological evolution affecting the structure of PINs
is gene
 duplication, either of e.g. a single gene or in
rare cases the whole genome, followed by divergence of the duplicates
through mutation and selection. A pair of genes that evolved from the
same ancestral gene is termed a paralogous gene pair, and the two proteins coded by such a gene pair
are referred to as a paralogous protein pair.
Remarkably, the whole genome of the yeast {\it S.~cerevisiae} is
believed to have duplicated approximately 100  million years
ago~{\cite{kellis::nature2004}}. Evolution ensuing duplication 
can silence duplicates. In fact, only about $10\%$ of the duplicated
genes are known to have been retained in {\it S.~cerevisiae}.
Obviously, just after a duplication event, paralogous proteins interact with exactly the same set of proteins.
However, evolution following the duplication event results in loss
of some shared interactions, or gain of
new interactions by one of the two paralogous proteins. Divergence following duplication
thus reduces interaction similarity of paralogues.
Early studies of  paralogues  in the yeast PIN
showed little or no interaction similarity~{\cite{wagner::mbe2001}}.
However, when paralogues resulting from the whole genome
duplication event (WGD) in {\it S.~cerevisiae} were studied separately, those paralogues were much more
likely than randomly chosen pairs to share at least one
neighbour~{\cite{musso::trends2007,troyanskaya::genetics2007}}, and were
significantly more likely to be co-clustered within the same protein
complex. This suggests that in  connected subgraphs much smaller than the entire
network WGD paralogues may be observed to have the same  neighbours. Indeed,
our findings reliably confirm this hypothesis.

\subsection{Summary}
In Section II, we define  twinness as a measure of node similarity within
local structures in networks, summarize the graph animal algorithm used to
estimate twinness, the Monte Carlo procedure used to construct null models,  and the experimental methods through which  the different 
PINs~{\cite{krogan::nature2006,gavin::nature2006,butland::nature2005}} we analyze were constructed.
In Section III, we present our results.
We measure twinness in connected, labeled subgraphs of size $n$ in
{\it E.~coli} and {\it S.~cerevisiae} PINs.
These are uniformly
sampled using the graph animal algorithm~{\cite{baskerville::arxives2007}}.  For each subgraph, pairs of
nodes that share identical subgraph neighbourhoods are identified. Twins are
divided into two types depending on whether (type ${\bf B}$) or not
(type ${\bf A}$) they are linked to each other.  For all PINs and  
for $n=4$ to $n=12$  twinness is (almost always) significantly
higher than expected from the null hypothesis. 
Our consistent observation for twinness for three different {\it S.~cerevisiae} PINs
speaks to its robustness as a structural feature of PINs.
Also, type ${\bf A}$ twinness is dominant for the prokaryote {\it E.~coli}, while type ${\bf B}$ twinness
is dominant for eukaryote {\it S.~cerevisiae}.  We observe that pairs of WGD paralogues in yeast are
grossly over-represented as twins compared to pairs chosen at random from
the network.   In Section IV we conclude with a discussion of some implications
of these results and give an outlook for
future directions.

\section{Definitions, Methods and Data} \label{methods} 
As is customary, we represent the connectivity of nodes in a graph $\mathcal{G}$ in
terms of an
adjacency matrix $\mathbf{A}$. For undirected, unweighted networks, each element
$A_{ij}=A_{ji}$ is either 0 -- if no link exists between node $i$ and
node $j$ --  or 1 otherwise.  Assume that node $i$ is contained within
a connected, labeled 
subgraph $\mathcal{H}$ of $\mathcal{G}$. The number of neighbors of $i$ in
$\mathcal H$ is its subgraph
degree $k_i^{\mathcal{H}}$. Two nodes $i,j$ are twins in 
$\mathcal{H}$ if they have identical neighbours in
$\mathcal{H}$, {\it i.e.}
\begin{equation} \label{eq:basic-condition}
        A_{i l} = A_{j l} \quad \forall l \in \mathcal{H}, \quad l\neq i, l\neq j \quad .
\end{equation}

\begin{figure}
\centering
\includegraphics[width=0.5\textwidth]{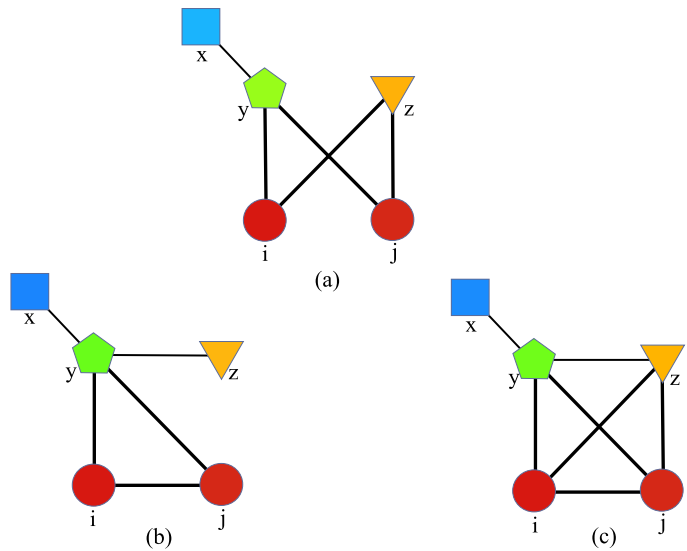}
\caption{Examples of twins in a connected, labeled subgraph of size $n=5$. Nodes $i$ and $j$ are twins.  (a) Type {\bf A} twins: defined as those satisfying $ A_{i l} = A_{j l} \quad \forall l \in \mathcal{H}$ and $k^\mathcal{H}_i, k^\mathcal{H}_j \geqslant 2 $.  (b) Type ${\bf B}$ twins: defined as those satisfying $A_{i l} = A_{j l} \mbox{  }l \neq i,j \mbox{  }\forall l \in \mathcal{H}$ with $k^\mathcal{H}_i, k^\mathcal{H}_j \geqslant 2$ and $A_{i j} = 1$.  (c) Type ${\bf B}'$ twins: defined as those satisfying $A_{i l} = A_{j l} \mbox{  }l \neq i,j \mbox{  } \forall l \in \mathcal{H}$ and $k^\mathcal{H}_i, k^\mathcal{H}_j \geqslant 3$ and $A_{i j} = 1$. All type ${\bf B}'$ twins are also type ${\bf B}$ twins.\label{fig:twin-defs-fig}}
\end{figure}

One consequence of Eq.~(\ref{eq:basic-condition}) is that $i$
and $j$ must have the same subgraph degree within
$\mathcal{H}$, or $k^{\mathcal{H}}_i=k^{\mathcal{H}}_j$.  To avoid the
trivial case of twins with subgraph degree equal to one  -- which occurs
when {\it e.g.} many nodes are connected to a hub --  we require that
$k_i^{\mathcal{H}}=k_j^{\mathcal{H}} \geq 2$. A pair of nodes $i,j$ with subgraph degrees larger
than two that satisfy Eq.(~\ref{eq:basic-condition}) are type ${\bf A}$ twins if $A_{ij}=0$ and type $\bf B$ twins if $A_{ij}=1$.  Biologically the distinction
between types ${\bf A}$ and ${\bf B}$ is motivated by the possibility that the ancestral protein of a paralogous pair was capable of homodimerization, {\it i.e.}
it could interact with itself. Immediately after its duplication, the two paralogous proteins interact with each other and subsequent
evolution can conserve this interaction.
Type ${\bf B}$ twins
with $k^{\mathcal H} \geq 3$ are further denoted as type ${\bf B}'$ twins. These definitions are illustrated in Fig.~\ref{fig:twin-defs-fig} for an $n=5$ connected, labeled subgraph.

Using the graph animal algorithm, explained below, we obtain an estimate of the total number of connected, labeled
$n$ node subgraphs, $\hat{H}^{n,{\bf X}}_{i j}$, 
in which the  pair $i,j$ are type ${\bf X}$ twins, with ${\bf X} = {\bf A} , ~\ {\bf B}, {\rm or}~\ {\bf B}'$.
We refer to $\hat{H}^{n,{\bf X}}_{i j}$ as the type ${\bf X}$ twinness of the pair $i,j$. Furthermore, we refer to the pair
$i,j$ as type ${\bf X}$ twins if $\hat{H}^{n,{\bf X}}_{i j}>0$, that is if  $i$ and $j$ appear as type $\bf X$ twins in at least one
connected, labeled $n$ node subgraph in the network.
Notice that a pair of nodes ($i,j$) cannot be both  type {\bf A} and type {\bf B} twins,
although the pair may appear as  twins in one subgraph but not in another. 
The total twinness for $n$ node subgraphs in a   network with $N$ nodes is given by
\begin{equation}\label{eq:exp-num-twins}
        \hat T^{{ \bf X}}_{n}  =\sum_{i,j}^N \hat H_{ij}^{n,{\bf X}} \quad
        .
\end{equation}
The carets in the above expressions
denote quantities that are estimates based on a sampling method. 
We also define the twinness per subgraph,
\begin{equation}\label{eq:exp-num-twins-density}
        \hat t^{{ \bf X}}_{n}  =\frac{\sum_{i,j}^N \hat H_{ij}^{n,{\bf X}}}{\hat H_n^N},
\end{equation}
to account for the dependence of twinness on the total number of connected,
labeled $n$ node subgraphs, ${\hat H_n^N}$, in the network. Twinness per subgraph enables comparisons between networks of different sizes and
also between the original networks and their randomized counterparts.

The graph animal algorithm~\cite{baskerville::arxives2007} makes it possible to 
obtain all of the estimates mentioned above.  The algorithm, a generalization of Leath algorithms for lattice animals~\cite{leath::prb1976}, 
grows connected subgraphs, and re-weights them to
achieve uniform sampling of all labeled, connected subgraphs.
Briefly, the graph animal algorithm proceeds as follows: (1) 
Pick a node at random from the network and mark it as
'infected'; (2) Add each neighbour of that node in the network to the boundary
list as `uninfected'; (3) While the number of  nodes in the subgraph is less than  
$n$ and the uninfected boundary list in not empty, do the following: 
Choose one uninfected boundary site and with probability $p$ mark it as 
infected, join it to the subgraph, and update the boundary list with the
neighbours of that node (which were not previously on the list) marked as
uninfected.  With probability $(1-p)$ mark the node as immune. Immune nodes can never be added to the subgraph. For more
details see~\cite{baskerville::arxives2007}.  The graph animal algorithm
enables uniform sampling of all labeled connected subgraphs of $n$ nodes
for a range of $n$.

Evaluating the statistical significance of the results requires a null model.  We use the standard~\cite{sneppen::science2002}
(although in some cases inadequate~\cite{baskerville::arxives2007})  null model  based on link exchange or rewiring. 
Then the results for connected labeled subgraphs can be compared with the
results for the same quantities in  an ensemble of graphs with the same degree sequence.  This also has the advantage of accommodating discrepancies in node degrees among different experimental constructions of the yeast PIN as indicated in Fig.~1.
To construct this null model,
 one randomly selects two links, $l_1$ and $l_2$ where, say,
 $l_1$ connects nodes $a$ and $b$ and $l_2$ connects $c$ and $d$. A link exchange move
 is  performed and the resulting links now connect $b$ to $d$
and $a$ to $c$. Exchanges  that would result in multiple or self-links are rejected. This process is  repeated many times, and any number of 
nominally independent
realizations of the null hypothesis can be constructed by repeating link
exchange moves.  

In this study we analyze  four recently constructed PINs, one for {\it E.~coli} and 
three for {\it S.~cerevisiae}. 
The {\it E.~coli} network was made 
by Butland {\it et al.} using experimental data obtained from High-Throughput Tandem Affinity 
Purification(HT-TAP), followed by MALDI-TOF MS and LC-MS to identify the purified proteins.  
The first {\it S.~cerevisiae} network was inferred
by Krogan {\it et al.} using a machine learning algorithm that 
utilized the MIPS database of known stable complexes as the gold 
standard upon which the algorithm was trained.  The raw experimental data for the Krogan {\it et al.} 
network was obtained using the same method as Butland
{\it et al.}  The second 
{\it S.~cerevisiae} network was inferred by Gavin {\it et al.} using a logarithmically weighted superposition 
of the spoke and matrix methods of link inference.  The raw experimental data was obtained
by Gavin {\it et al.} using HT-TAP, followed by MALDI-TOF MS.  Data for the third {\it S.~cerevisiae} network was
collected by  Batada {\it et al.} 
through a comprehensive literature review of all known sources of 
Protein-Protein Interaction data. The network assembled by Batada {\it et al.} included only those interactions that were confirmed 
by several published sources, or were verified  by a highly accurate method (such as Western Blot).  The data for all these networks can be found in the on-line supplementary materials associated with Refs.~{\cite{krogan::nature2006,gavin::nature2006,batada::plos2006,butland::nature2005}}.  Table~\ref{tab:netstats} summarizes various global properties of these networks, and their degree distributions are shown in
Fig.~\ref{fig:degdist}.

\begin{table}
\centering
\begin{tabular}{|l|l|l|l|l|}
\hline
{\bf Source}&{\bf Organism}& $N$&{ $L$}&$\langle k\rangle$\\
\hline
Butland {\it et al.}&{\it E.~coli}&230&695&6.0\\
Krogan {\it et al.}&{\it S.~cerevisiae}&2559&7031&5.5\\
Gavin {\it et al.}&{\it S.~cerevisiae}&1374&6833&9.9\\
Batada {\it et al.}&{\it S.~cerevisiae}&2752&9097&6.6\\
\hline
\end{tabular}
\caption{Global properties of the protein interaction
 networks studied. The quantity $N$
is the number of nodes in the network, $L$ is the number of links in the
network, and $\langle k \rangle$ is the average degree of a node in the
network.\label{tab:netstats}}
\end{table}

The data set of Whole Genome Duplication (WGD) paralogues for {\it S.~cerevisiae}
was  produced by Kellis {\it et al.}~\cite{kellis::nature2004} and is found in the 
Supplementary Material of that publication.  By sequencing and analyzing the genome of {\it Kluyveromyces waltii} --
a related yeast species --  these authors
were able to identify paralogous genes resulting from a whole gene duplication
event that occurred shortly after the two species diverged.  The Kellis {\it
et al.} data set was translated~\cite{musso::trends2007} so that it could be applied
to the Krogan {\it et al.} data.

\section{Results}\label{results}

\begin{figure}
\centering
\includegraphics[width=0.6\textwidth]{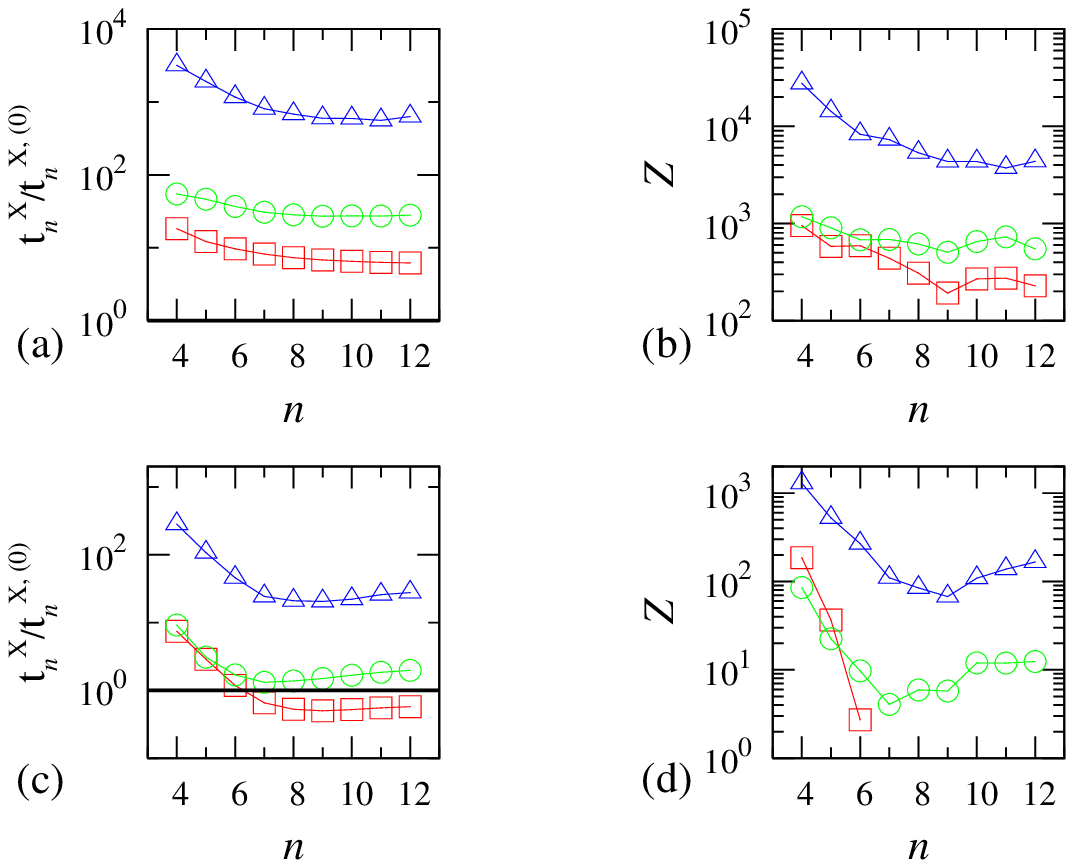}
\includegraphics[width=0.6\textwidth]{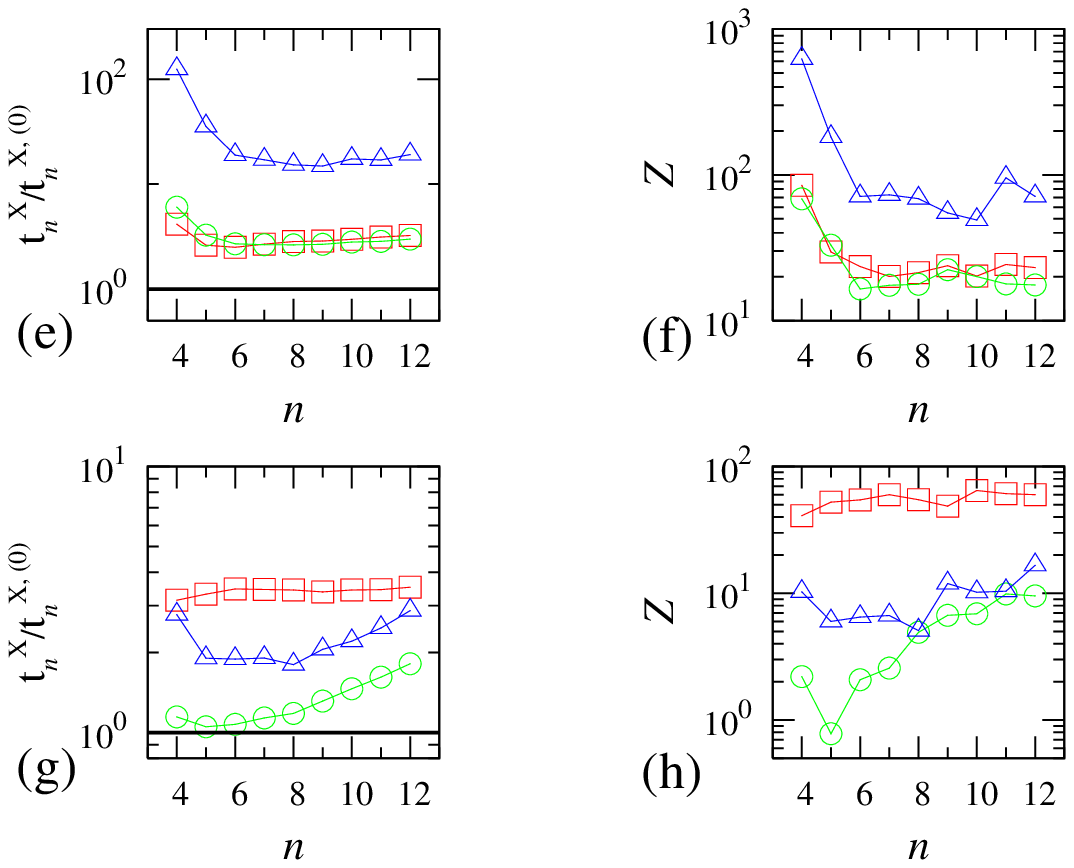}
\caption{ Ratios of type ${\bf A}$ , ${\bf B}$ and ${\bf B}'$ 
twins per  subgraph in the original PIN, ${\hat t}^{\bf X}_n$, to twins per subgraph in the corresponding
null model, ${\hat t}^{{\bf X},(0)}_n$, as a function of $n$,
accompanied by $Z$ scores for  type ${\bf A}$ , ${\bf B}$ and  ${\bf B}'$ 
twins per subgraph as a function of $n$. (a) and (b) are for the data set 
of Gavin {\it et al.}, (c) and (d) for Batada {\it et al.}, 
(e) and (f) for Krogan {\it et al.}, and (g) and (h) for the {\it E.~coli} data set. 
Type ${\bf A}$  is represented by squares, type ${\bf B}$ by circles and 
type ${\bf B}'$ by triangles for each data set.
All twin ratios, except type ${\bf
A}$  in the Batada {\it et al.} data, are approximately independent
of $n$ at larger $n$ and significantly over-abundant, with $Z$ scores
ranging up to $\approx 10,000$.
\label{fig:numberandzeds}}
\end{figure}

The first result of our analysis is a much higher twinness in the PINs than expected.
As demonstrated in Fig.~\ref{fig:numberandzeds}, in all cases
twinness per subgraph is higher than in the null model -- 
 except for type
${\bf A}$ twins in the Batada {\it et al.} PIN.   
This is in stark contrast 
with the result obtained for twinness in the limit $n \to N$, where no twins of 
any type appear in either the original networks or in their randomized versions. In other words, considering the entire PIN, no two nodes of degree greater than two share 
all of their neighbours.

The statistical significance of the results for twinness
 is also demonstrated in Fig.~\ref{fig:numberandzeds},
which shows the $Z$ scores \footnote{$Z$ score for a measurement $x$ relative to an ensemble of null model measurements $\{x^{(0)}\}$ is:
$Z=\frac{x-\langle x^{0} \rangle}{\sigma^{(0)}}$, where $\sigma^{(0)}$ is the standard deviation of $x$ within the null ensemble.} for twinness per subgraph in the original
PINs compared to their randomized counterparts. 
This data is also plotted on log-linear axes.  Almost all $Z$ scores
in Fig.~\ref{fig:numberandzeds} are greater than 1, and some, particularly for type ${\bf B}'$ 
twins are several orders of magnitude larger.  These extremely high $Z$ scores indicate that
rewired networks are not the optimal null model for PINs (see Refs.~\cite{baskerville::arxives2007,newman::siam2003}.)
Nevertheless the values obtained both for twin ratios and their corresponding $Z$ scores
are relatively stable for $n\gtrsim 8$.  This indicates that a precise choice
of $n$ is not necessary.

\begin{figure}
\centering
\includegraphics[width=0.7\textwidth]{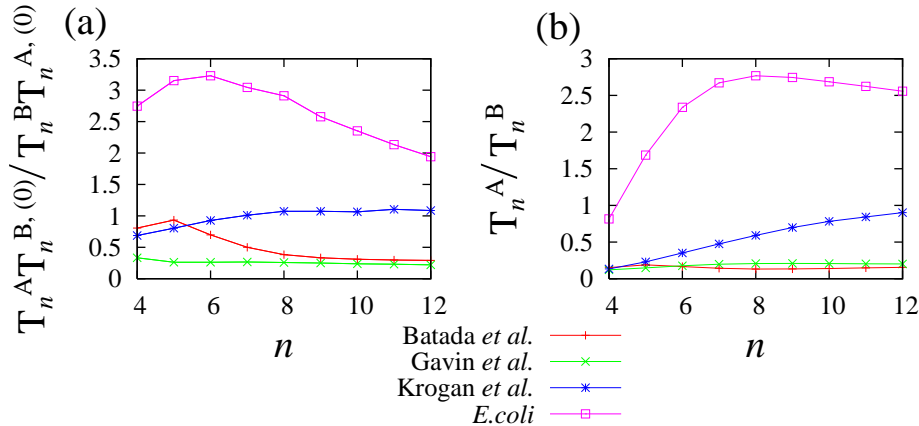}
\caption{(a) Ratio of type ${\bf A}$ total twinness to type ${\bf B}$
total twinness in a connected subgraph of size $n$, normalized by the average of the corresponding twinness for the randomized networks. (b) Ratio of type ${\bf A}$ total twinness to type ${\bf B}$
total twinness in connected subgraphs of size $n$.  Error bars are not shown due to the fact
that they would be smaller than the symbols.  Errors include both the
uncertainty associated with sampling and (for (a)) the variance in the values
for the ensemble of randomized networks. Note that in both cases
type ${\bf A}$ twins dominate for {\it E.~coli} while type ${\bf B}$ twins dominate
for {\it S.~cerevisiae}.\label{fig:AB}}
\end{figure}

The second result of our analysis is a statistically significant difference  between the 
ratio of type ${\bf A}$ to type ${\bf B}$ twinness in {\it S.~cerevisiae}
PINs compared to the   {\it E.~coli} PIN. 
In Fig.~\ref{fig:AB}b we see that the ratio of type ${\bf A}$ twinness to  
type ${\bf B}$ twinness is much higher for the {\it E.~coli} PIN than for all three {\it S.~cerevisiae} 
PINs.   In Fig.~\ref{fig:AB}a we normalize this ratio by the same ratio for the respective ensemble of 
randomized networks.  This accounts for the differences in  degree sequence among the yeast PINs.
The difference between {\it E.~coli} and {\it S.~cerevisiae} persists after this normalization.  
It is particularly important that the Gavin {\it et al.} 
and Krogan {\it et al.} {\it S.~cerevisiae} PINs are easily distinguished from the
Butland {\it et al.} {\it E.~coli} PIN because the experimental data for all three of these 
networks was obtained using the TAP-MS method.  The systematic decline for  {\it E.~coli} in Fig. \ref{fig:AB} might be a finite size effect. In fact, this 
PIN is an order of magnitude smaller than the {\it S.~cerevisiae} PINs (see Table~\ref{tab:netstats}). 

\begin{figure}
\centering
\includegraphics[width=0.7\textwidth]{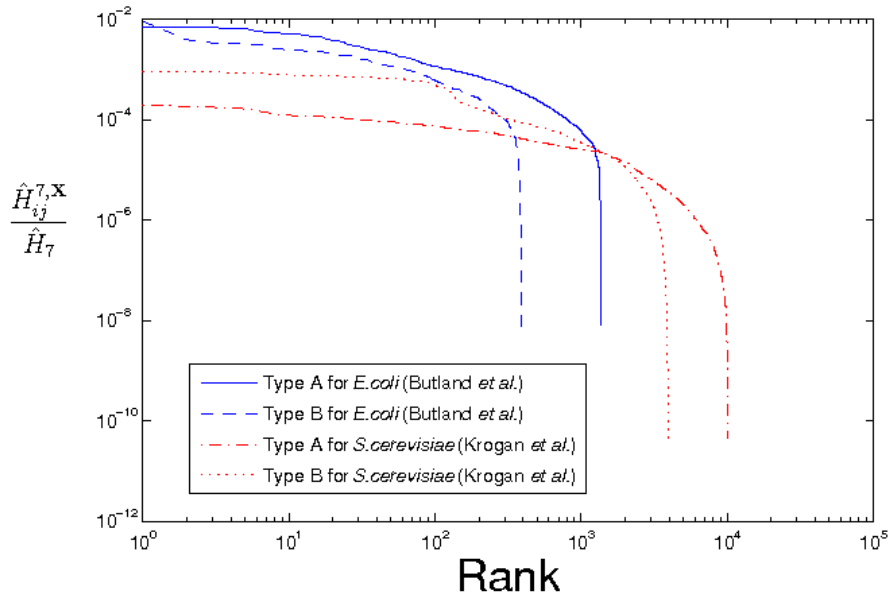}
\caption{Zipf plot for pairwise twinness in 7 node subgraphs. 
Each pair $i,j$ is ranked according to its twinness, $\hat{H}_{i,j}^{7,{\bf X}}$, and then plotted
against its relative rank. Note that the frequencies for type {\bf A} twins
are greater than those for type ${\bf B}$ twins for ${\it E.~coli}$ while the
opposite is the case for {\it S.~cerevisiae}. 
The same trend is observed for the other two 
(Batada \textit{et al.}, Gavin \textit{et al.}) \textit{S.~cerevisiae} data sets, as well as other 
subgraph sizes. \label{fig:zipf}}
\end{figure}

Fig.~\ref{fig:zipf} is a Zipf plot of twinness for each pair of nodes.
We rank each pair $i,j$ according to its twinness, $\hat{H}_{i,j}^{n,{\bf X}}$, and then plot this twinness value against its relative rank.
Pairwise twinness and ranks for Fig.~\ref{fig:zipf} were evaluated for $n=7$ node subgraphs. 
To facilitate the comparison of different PINs, 
we divide each pair's twinness
by $\hat{H}_7$, the estimated total number of seven node subgraphs in the respective networks. 
Zipf plots for other $n$'s and data sets  are qualitatively similar to those shown for the corresponding organism.

The difference between the prokaryote ({\it E.~coli}) and the eukaryote ({\it S.~cerevisiae}) PINs  
exhibited in Fig.~\ref{fig:AB} could be due to a higher number  of 
type ${\bf A}$ twins than type ${\bf B}$ twins in {\it E.~coli} -- and the opposite in {\it S.~cerevisiae}. 
Fig.~\ref{fig:zipf} shows this is not the case, as for both the Krogan {\it et al.} {\it S.~cerevisiae} PIN and the
Butland {\it et al.} {\it E.~coli} PIN the number of type ${\bf A}$ twins is roughly the same order of magnitude as the 
the number of type ${\bf B}$ twins.  For {\it E.~coli}  the typical twinness of a type ${\bf A}$ twin
is larger than the typical twinness of a type ${\bf B}$ twin, while the opposite is true for {\it S.~cerevisiae}.

\begin{table}

\centering
{\footnotesize
\begin{tabular}{|l|l|l|l|l|l|l|l|}
\hline
{\bf Type}&{\bf Measure}&{\bf $n=5$}&{\bf $n=6$}&{\bf $n=7$}&{\bf $n=8$}&{\bf $n=9$}&{\bf $n=10$}\\
\hline\hline
${\bf A}$&Number of twins in set&11&11&11&11&11&11\\
&of WGD paralogous pairs.&&&&&&\\
\cline{2-8}
&Same as above for an&$0.36$&$0.36$&$0.36$&$0.36$&$0.36$&$0.36$\\
&ensemble of random samples.&&&&&&\\
\cline{2-8}
&$Z$ scores for total subgraphs&&&&&&\\
&in which paralogues appear as&$4.84$&$2.45$&$1.49$&$1.08$&$0.89$&$0.75$\\
&twins.&&&&&&\\
\hline\hline
${\bf B}$&Number of twins in set&17&17&17&17&17&17\\
&of WGD paralogous pairs.&&&&&&\\
\cline{2-8}
&Same as above for an&$0.14$&$0.14$&$0.14$&$0.14$&$0.14$&$0.14$\\
&ensemble of random samples.&&&&&&\\
\cline{2-8}
&$Z$ scores for total subgraphs&&&&&&\\
&in which paralogues appear as&$30.72$&$22.85$&$19.04$&$17.25$&$17.42$&$17.68$\\
&twins.&&&&&&\\
\hline\hline
${\bf B}'$&Number of twins in set&12&12&12&12&12&12\\
&of WGD paralogous pairs.&&&&&&\\
\cline{2-8}
&Same as above for an&$0.11$&$0.11$&$0.11$&$0.11$&$0.11$&$0.11$\\
&ensemble of random samples.&&&&&&\\
\cline{2-8}
&$Z$ scores for total subgraphs&&&&&&\\
&in which paralogues appear as&$25.22$&$30.96$&$27.53$&$23.92$&$21.49$&$20.89$\\
&twins.&&&&&&\\
\hline
\end{tabular}
}
\caption{Results for the set of 117 WGD paralogous pairs found in the {\it S.~cerevisiae} PIN of Krogan {\it et al.}  
The first two rows for each type compare the number of twins found in the WGD paralogue set to 
the average number of twins expected when choosing 117 pairs at random from
the network. 
WGD paralogues are twins $\approx 100$ times more often than pairs chosen at random.  The third row is the $Z$  score
 for the total twinness of the paralogues, compared to the total twinness of an ensemble of sets of 117 randomly selected pairs of nodes.
The $Z$ scores establish the statistical significance of the observed correlation of WGD paralogues with type ${\bf B}$ twinness.\label{tab:WGD}}
\end{table}

We now turn our attention to our third result, examining relationships
between twinness and the evolutionary histories of PINs.
Specifically, we focus on the
 set of paralogous proteins resulting from the whole genome duplication
event in an ancestor of {\it S.~cerevisiae}~\cite{kellis::nature2004}.
We started with the set of 457 WGD paralogous 
pairs identified by Kellis {\it et al.}~\cite{kellis::nature2004}.
We then searched for the WGD paralogues in the Krogan {\it et al.} {\it S.~cerevisiae} PIN, 
and identified 117 WGD paralogues for which both member proteins were present in the PIN.  The twinness 
of these 117 WGD paralogues was then measured, and a summary of this analysis is presented in Table \ref{tab:WGD}.

Our results indicate a strong correlation between WGD paralogous pairs and twins.  We count the number 
of type ${\bf A}$, ${\bf B}$, and ${\bf B}'$ twins in the set of 117 paralogues and compared this to 
the same quantity averaged over many samples obtained by choosing 117 pairs
at random from the PIN.  The WGD paralogue set contains many more 
twins, of all types, than expected from a random set of node pairs.  For type ${\bf A}$ we find 11 WGD paralogues 
that appear as twins, compared to an average of 0.36 for random samples of 117 pairs.  Likewise for types
${\bf B}$ and ${\bf B}'$ we find 17 compared to 0.14 and 12 compared to 0.11, respectively. For all types 
the number of WGD paralogue twins is at least one order of magnitude greater than expected, and 
in the cases of ${\bf B}$ and ${\bf B}'$, two orders of magnitude.  
In addition, WGD paralogues have a higher twinness (for all types) than  pairs of 117 nodes chosen at random. 
However, when $Z$ scores are calculated, this 
effect remains statistically significant only for type ${\bf B}$ and ${\bf B}'$.

To guide the reader through Table~\ref{tab:WGD} let us look at the case of type ${\bf A}$ twins for
$n=5$ subgraphs.  The entry of the first row indicates that 11 of the 117
WGD paralogues appear in at least
one $n=5$ subgraph as type ${\bf A}$ twins, while the second row indicates that the number of type ${\bf A}$ twins expected 
in a sample of 117 randomly selected pairs is 0.36.  The entry in the third row is the $Z$ score for the total
twinness, $\hat T^{{ \bf A}}_{5}$, summed only over the 117 WGD paralogues, compared to the total twinness of a random sample of 117 pairs.
It should be noted that in the case of type ${\bf A}$
twins the $Z$ score falls below 1 for large subgraphs in the Krogan {\it
et al.} PIN.  This is due to the fact that although there is large 
number of twins among the WGD paralogues ($11 \gg 0.36$), their actual twinness values are low.  From the results presented in Table~\ref{tab:WGD} it is apparent that  twinness (for type ${\bf B}$) correlates 
strongly with WGD paralogues.  This suggests that further refinements of our methods could result 
in a technique capable of identifying paralogous genes based on the network properties of PINs.

\section{Discussion}\label{discussion}
Early interest in the topological structure of protein interaction networks
(PINs) was driven in part by the observation
of a scale-free degree distribution~\cite{barabasi::nature2001}.
 Since then various other
topological properties of PINs have been studied, such as their disassortativity~\cite{sneppen::science2002},
clustering and over-abundance of 
small subgraphs or motifs~\cite{baskerville::arxives2007}.
Many of these structural properties have failed the test of robustness 
presented by newer higher confidence data~\cite{batada::plos2006,coulomb::prsb2005} -- as also indicated in Fig.~1.
Other properties that may be correlated with underlying biological criteria, such as protein 
complexes~\cite{spirin::pnas2003}, have proved to be more robust~\cite{krogan::nature2006,gavin::nature2006},
although some~\cite{wang::plos2007} have argued that protein complexes
identified using clustering techniques on PINs lack biological significance.

In this paper we have proposed and studied a new structural property of networks that is not based on clustering algorithms or the identification
of complexes --  namely the occurrence of
twin nodes in connected subgraphs. Our observation that PINs exhibit significantly
higher twinness than randomized
networks is robust for three different {\it S.~cerevisiae} PINs 
despite the empirical disparity in the global characteristics of these networks.
A further robust result is the marked difference between type ${\bf A}$ and
type ${\bf B}$ twinness separating the {\it E.~coli} PIN from all three {\it
S.~Cerevesiae} PINs studied.  Based on the
gross over-representation of
WGD paralogues as type ${\bf B}$ twins, we propose that our measure is associated with the biological phenomenon of
gene duplication.

To show that the PINs studied possessed statistically significantly higher twinness than expected, a null model was used.  The extremely high $Z$ scores observed in our analysis (see Fig.~\ref{fig:numberandzeds}) makes it clear that this null model is not optimal for PINs.  The commonly held picture of the global structure of a PIN is that it consists of a number of dense clusters of nodes, that are in turn loosely connected to each other.  It is well documented that rewiring significantly reduces the frequency of dense subgraphs~\cite{baskerville::arxives2007,newman::siam2003}.  Thus, due do the fact that rewiring destroys the characteristic dense clusters mentioned above, it is unlikely that a null model based on purely random rewiring is the best null model for these networks.

This specifically applies to our results in the following manner.  A subgraph with many links will tend to have a large number of types ${\bf B}$ and ${\bf B}'$ twin pairs. (For instance, all pairs are twins in a complete graph.)  However, in the rewired network, there are many fewer dense subgraphs and as such, many fewer types ${\bf B}$ and ${\bf B}'$ twin pairs are observed.  As well the over-represented motifs of the networks we analyzed are known to be dense.  These points offer a potential explanation for the high levels of types ${\bf B}$ and ${\bf B}'$ twinness observed.  While this probably affects our first result (over-representation of twins in PINs), it could also affect our second result on the difference in levels of type ${\bf A}$ and type ${\bf B}$ twinness between {\it E.~coli} and {\it S.~cerevisiae} -- as discussed below.  However, the short comings of this null model could not impact our third result, the observed correlation between type ${\bf B}$ twinness and WGD paralogues.

The difference in the ratio of type ${\bf A}$ to type ${\bf B}$ twinness between the prokaryote {\it E.~coli} and eukaryote {\it S.~cerevisiae} could be explained if the {\it E.~coli} network exhibited a fundamentally different type of clustering than the three {\it S.~cerevisiae} networks.  This is suggested by the results of ~\cite{baskerville::arxives2007}, as the extended motifs of the {\it E.~coli} network tend to be bipartite subgraphs (which contain many type ${\bf A}$ twins), while the extended motifs of the {\it S.~cerevisiae} network are complete, or nearly complete subgraphs (which contain many type ${\bf B}$ twins).  However, the connection between motifs and twins is not as trivial as it might first appear.  Motifs are over-represented, but do not necessarily comprise the bulk of subgraphs.  Another possible explanation for the ratio difference comes from biological considerations.  Consider a twin pair of proteins descended from the same ancestral protein capable of homodimerization.  It would follow that the twin proteins would be candidates to be type ${\bf B}$ twins if the interaction arising from the homodimerization of the ancestral protein was conserved, where they would be candidates to be type ${\bf A}$ twins if that interaction was lost.  Thus our results for the type ${\bf A}$ to type ${\bf B}$ twinness ratios in {\it E.~coli} and {\it S.~cerevisiae} could arise from a) a fundamental difference in the frequency of homodimerizing proteins between the two organism; b) differing rates of conservation of an ancestral homodimerization interaction between the two organisms or c) for reasons not related to homodimerization, such as different types of clustering.  It is clear that to determine which of these cases is most likely a more sophisticated null model would be useful, particularly if it could reproduce the different types of clustering, and extended motifs found in the {\it E.~coli} and {\it S.~cerevisiae} PINs.

Turning our attention to our third result we note that paralogous pairs resulting from the whole genome duplication event in {\it S.~cerevisiae} have been reported to be more likely to share at least one neighbour~\cite{troyanskaya::genetics2007,musso::trends2007}. While this is consistent with our result for twinness of WGD paralogues, no two nodes with degrees larger than two share all their neighbours.  Our result is also consistent with the observation that WGD paralogues are separated by  a shorter minimum-edge distance~\cite{musso::trends2007} and should thus appear in subgraphs of sizes that we have considered.  Our analysis suggests that ancestral relationships between proteins can perhaps be uncovered using twinness in subgraphs, with the caveat that  neither are all twins WGD paralogues, nor are all WGD paralogues  twins.  Furthermore it is possible that small scale duplicate proteins (such as those identified in~\cite{troyanskaya::genetics2007}) will also appear as twins in the PIN.  However, as discussed earlier, the high link density of motifs can be responsible for high twinness between nodes. Disentangling the effects of clusters from the effects of paralogous relationships is a subject for future work.

Moving on to directions not directly related to our results, we contend that analyses of connected subgraphs can shed light on the inference methods used to construct PINs. Recent TAP based experiments purify groups of proteins that are directly or indirectly bound to a bait protein. This protein group data is then converted into an interaction network using various inference methods~\cite{bader::naturebiotech2002,krogan::nature2006,gavin::nature2006}. Proteins that are purified as a group should lie close to each other in the inferred network, and thus be captured in our subgraphs. The criteria that were used to identify pairwise interactions from the protein group data should have a direct effect on the twinness of proteins. In fact, the gross differences between the PIN of Krogan {\it et al.} and that of Gavin {\it et al.}, both of which are based on TAP experiments, might be due to the different inference methods used~\cite{uetz::genbio2006}.  We have not pursued  questions related to inference methods in this work, but it does present a future direction of inquiry.

Finally, we point out that apart from biological motivations, twinness between nodes
can also be used as a node similarity measure for other real-world networks.
Node similarity based on the intersection of the
neighbour sets has been used to study similarity between
documents~{\cite{kessler::AD1963,small::jasis1973}}, and has been extended
to iterative definitions of similarity~\cite{jeh::acm2002,newman::pre2006}.
The basic premise for these similarity measures, as for twinness, is 
that if the links between nodes are derived from their
function, then the functional similarity between nodes can be inferred
from the topological structure of the PIN. The method presented here
is not iterative, but focuses on node similarity within local network  structure,
which is arguably of functional relevance.

\bibliography{twins}
\end{document}